\documentclass[prd,onecolumn,nofootinbib,showpacs,showkeys]{revtex4}
\newcommand\gothfamily{\usefont{U}{ygoth}{m}{n}}
\DeclareTextFontCommand{\textgoth}{\gothfamily}

\begin{document}

\title{VARIATIONAL FORMULATION OF EISENHART'S UNIFIED THEORY}

\author{{\bf Nikodem J. Pop\l awski}}

\affiliation{Department of Physics, Indiana University, Swain Hall West, 727 East Third Street, Bloomington, IN 47405, USA}
\email{nipoplaw@indiana.edu}

\noindent
{\em International Journal of Modern Physics A}\\
Vol. {\bf 24}, Nos. 20 \& 21 (2009) 3975--3984\\
\copyright\,World Scientific Publishing Co.
\vspace{0.4in}

\begin{abstract}
Eisenhart's classical unified field theory is based on a non-Riemannian affine connection related to the covariant derivative of the electromagnetic field tensor.
The sourceless field equations of this theory arise from vanishing of the torsion trace and the symmetrized Ricci tensor.
We formulate Eisenhart's theory from the metric-affine variational principle.
In this formulation, a Lagrange multiplier constraining the torsion becomes the source for the Maxwell equations.
\end{abstract}

\pacs{04.20.Fy, 04.50.Kd}
\keywords{Unified field theory; metric-affine gravity.}
\maketitle

\section{Introduction}

In general relativity, the electromagnetic field and its sources are considered to be on the side of the matter tensor in the Einstein field equations, acting as sources of the gravitational field.
The geometry of general relativity is that of a four-dimensional Riemannian manifold, equipped with a symmetric metric-tensor field and an affine connection that is torsionless and metric compatible.
In classical unified field theories, the electromagnetic field obtains the same geometrical status as the gravitational field~\cite{Goe}.
In order to combine gravitation and electromagnetism on the classical level within a geometrical theory we must modify some postulates of general relativity, resulting in a non-Riemannian geometry~\cite{Eis0}.

The most known attempts of creating a unified field theory include: Weyl's conformal geometry~\cite{Weyl1,Weyl2}, Kaluza's five-dimensional theory~\cite{Kaluza1,Kaluza2,Kaluza3}, and the Einstein-Straus-Schr\"{o}dinger nonsymmetric field theory~\cite{Schrod1,Schrod2,Schrod3,Schr}.
Weyl relaxed the postulate of metric compatibility of the affine connection, obtaining a unified theory, where electromagnetic gauge transformation was related to conformal transformation of the metric.
In Kaluza's theory, electromagnetic potentials were represented in a five-dimensional metric, and the Lagrangian proportional to the five-dimensional Ricci scalar yielded the Einstein-Maxwell field equations and the Lorentz equation of motion.
Relaxing the postulate of the symmetry of the affine connection~\cite{Car} and metric tensor resulted in the Einstein-Straus-Schr\"{o}dinger theory that related the electromagnetic field to the skewsymmetric part of the metric tensor.
Although all three theories turned out to be unphysical~\cite{Lord1,Lord2}, Weyl's theory introduced the concept of gauge invariance that led to modern particle physics based on non-Abelian SU(n) fields, while Kaluza's theory inspired later models of spacetime with extra dimensions.

Later unified theories related the electromagnetic field to the non-Riemannian part of the affine connection rather than to the metric.
The connection generalizes an ordinary derivative of a vector into a coordinate-covariant derivative, while the electromagnetic potential generalizes it into a $U(1)$-covariant derivative.
Associating the electromagnetic potential with the connection seems more natural because both objects have the same purpose: to preserve the correct transformation properties under certain symmetries.
The electromagnetic potential can be identified with the trace of the nonmetricity tensor (accordingly, the electromagnetic field tensor is represented by the segmental curvature tensor)~\cite{FKu,PO}, or with the trace of the torsion tensor~\cite{Ham}.
In these theories, the Einstein-Maxwell field equations were generated from varying the action with respect to the connection~\cite{FKu}, both the connection and metric~\cite{PO}, or both the torsion and metric~\cite{Ham}.

Eisenhart showed~\cite{Eis1} that one can unify the Maxwell field equations and the Lorentz equation of motion with gravity inside a purely geometrical theory.
In Eisenhart's unified field theory, the torsion tensor is related to the Riemannian covariant derivative of the electromagnetic field tensor, and the torsion trace vanishes.
This theory was explored further in Refs.~\cite{Eis2,Eis3,Eis4}, however, as in Ref.~\cite{Eis1}, the field equations were postulated outside a variational principle.
In this paper, we formulate Eisenhart's unified theory as a variational theory.
Because the affine connection is independent of the metric, we use the metric-affine formulation of gravity~\cite{MAE1,MAE2,MA1,MA2,MA3,MA4,MA5,MA6}, in which both the metric tensor and affine connection are variables (gravitational potentials) with respect to which the total action is varied.
This formulation is dynamically equivalent to the purely metric Einstein-Hilbert formulation~\cite{FKe}.
We find and discuss the metric-affine Lagrangian that generates Eisenhart's field equations from varying the connection.
We include the condition that the torsion trace be zero as a Lagrange-multiplier constraint in the action.
This vector turns out to be the source current for the Maxwell equations.

\section{Eisenhart's Unified Field Theory}

Eisenhart's original theory~\cite{Eis1} is based on the affine connection $\Gamma^{\,\,\rho}_{\mu\,\nu}$ that is not metric compatible, depending on the symmetric metric tensor $g_{\mu\nu}$ via the Christoffel symbols $\{^{\,\,\rho}_{\mu\,\nu}\}=\frac{1}{2}g^{\rho\lambda}(g_{\nu\lambda,\mu}+g_{\mu\lambda,\nu}-g_{\mu\nu,\lambda})$, and the electromagnetic field tensor $F_{\mu\nu}=A_{\nu,\mu}-A_{\mu,\nu}$:
\begin{equation}
\Gamma^{\,\,\rho}_{\mu\,\nu}=\{^{\,\,\rho}_{\mu\,\nu}\}+kF_{\mu\nu:\sigma}g^{\rho\sigma},
\label{con}
\end{equation}
where the colon denotes the Riemannian covariant differentiation with respect to the Levi-Civita connection $\{^{\,\,\rho}_{\mu\,\nu}\}$, and the physical constant $k$ guarantees the correct dimension.\footnote{
Eisenhart chose the units such that $k=1$.
}
Accordingly, the symmetric part of the connection is equal to the Christoffel symbols, while its skewsymmetric part, the Cartan torsion tensor $S^\rho_{\phantom{\rho}\mu\nu}=\Gamma^{\,\,\,\,\rho}_{[\mu\,\nu]}$, contains the electromagnetic field:
\begin{equation}
S^\rho_{\phantom{\rho}\mu\nu}=kF_{\mu\nu:\sigma}g^{\rho\sigma}.
\label{tor}
\end{equation}
The trace of the torsion tensor, the torsion vector $S_\mu=S^\nu_{\phantom{\nu}\mu\nu}$, is thus
\begin{equation}
S_\mu=kF_{\mu\phantom{\nu}:\nu}^{\phantom{\mu}\nu}.
\label{trace}
\end{equation}
Imposing vanishing of the torsion vector,
\begin{equation}
S_\mu=0,
\label{Max}
\end{equation}
yields the sourceless Maxwell equations: $F_{\mu\phantom{\nu}:\nu}^{\phantom{\mu}\nu}=0$.
The nonmetricity tensor $N_{\mu\nu\rho}=g_{\mu\nu;\rho}$~\cite{Eis0,Scho},\footnote{
The affine connection is completely determined by the Christoffel symbols, torsion and nonmetricity~\cite{MA1,MA2}: $\Gamma^{\,\,\rho}_{\mu\,\nu}=\{^{\,\,\rho}_{\mu\,\nu}\}+S^\rho_{\phantom{\rho}\mu\nu}+2S_{(\mu\nu)}^{\phantom{(\mu\nu)}\rho}+\frac{1}{2}N_{\mu\nu}^{\phantom{\mu\nu}\rho}-N^\rho_{\phantom{\rho}(\mu\nu)}$.
}
where the semicolon denotes the covariant differentiation with respect to $\Gamma^{\,\,\rho}_{\mu\,\nu}$, is given by
\begin{equation}
N_{\mu\nu\rho}=k(F_{\rho\mu:\nu}+F_{\rho\nu:\mu}).
\label{non}
\end{equation}

Eisenhart's affine connection~(\ref{con}) gives the curvature tensor $R^\rho_{\phantom{\rho}\mu\sigma\nu}=\Gamma^{\,\,\rho}_{\mu\,\nu,\sigma}-\Gamma^{\,\,\rho}_{\mu\,\sigma,\nu}+\Gamma^{\,\,\kappa}_{\mu\,\nu}\Gamma^{\,\,\rho}_{\kappa\,\sigma}-\Gamma^{\,\,\kappa}_{\mu\,\sigma}\Gamma^{\,\,\rho}_{\kappa\,\nu}$~\cite{Eis0,Scho}:
\begin{equation}
R^\rho_{\phantom{\rho}\mu\sigma\nu}=K^\rho_{\phantom{\rho}\mu\sigma\nu}+k(F^{\phantom{\mu\nu}:\rho}_{\mu\nu\phantom{:\rho}\sigma}-F^{\phantom{\mu\sigma}:\rho}_{\mu\sigma\phantom{:\rho}\nu})+k^2(F^{\phantom{\mu\nu}:\kappa}_{\mu\nu}F^{\phantom{\kappa\sigma}:\rho}_{\kappa\sigma}-F^{\phantom{\mu\sigma}:\kappa}_{\mu\sigma}F^{\phantom{\kappa\nu}:\rho}_{\kappa\nu}),
\label{curv}
\end{equation}
the Ricci tensor $R_{\mu\nu}=R^\rho_{\phantom{\rho}\mu\rho\nu}$:
\begin{equation}
R_{\mu\nu}=K_{\mu\nu}+kF_{\mu\nu:\rho}^{\phantom{\mu\nu:\rho}\rho}+k^2F_{\mu\phantom{\rho}:\sigma}^{\phantom{\mu}\rho}F_{\nu\phantom{\sigma}:\rho}^{\phantom{\nu}\sigma},
\label{Ric}
\end{equation}
and the curvature scalar $R=R_{\mu\nu}g^{\mu\nu}$:
\begin{equation}
R=K+k^2F_{\mu\phantom{\nu}:\rho}^{\phantom{\mu}\nu}F^{\mu\rho}_{\phantom{\mu\rho}:\nu},
\label{sca}
\end{equation}
where $K^\rho_{\phantom{\rho}\mu\sigma\nu}$, $K_{\mu\nu}$ and $K$ are the corresponding Riemannian tensors, constructed from $\{^{\,\,\rho}_{\mu\,\nu}\}$ instead of $\Gamma^{\,\,\rho}_{\mu\,\nu}$.
The skewsymmetric tensor of homothetic curvature $Q_{\mu\nu}=R^\rho_{\phantom{\rho}\rho\mu\nu}=\Gamma^{\,\,\rho}_{\rho\,\nu,\mu}-\Gamma^{\,\,\rho}_{\rho\,\mu,\nu}$, which is proportional to the curl of the trace of the nonmetricity tensor: $Q_{\mu\nu}=-\frac{1}{2}(N^\rho_{\phantom{\rho}\rho\nu,\mu}-N^\rho_{\phantom{\rho}\rho\mu,\nu})$, vanishes because of Eqs.~(\ref{Max}) and~(\ref{non}).
As the gravitational field equation in vacuum, Eisenhart chose vanishing of the symmetrized Ricci tensor~\cite{Eis0}:
\begin{equation}
R_{(\mu\nu)}=0,
\label{Ein}
\end{equation}
or
\begin{equation}
K_{\mu\nu}=-k^2F_{\mu\phantom{\rho}:\sigma}^{\phantom{\mu}\rho}F_{\nu\phantom{\sigma}:\rho}^{\phantom{\nu}\sigma},
\label{Eis}
\end{equation}
which differs from Einstein's $K_{\mu\nu}=0$.
Equations~(\ref{con}),~(\ref{Max}) and~(\ref{Ein}) are the field equations of this theory, giving the metric tensor as a function of the coordinates.

The Lorentz equation of motion in Eisenhart's theory results from the condition~\cite{Eis0}
\begin{equation}
g_{\mu\nu}u^\mu u^\nu=\mbox{const},
\label{norm}
\end{equation}
where $u^\mu=\frac{dx^\mu}{ds}$ is the four-velocity vector tangent to a world line parametrized by $s$, $x^\mu=x^\mu(s)$.
Differentiating Eq.~(\ref{norm}) with respect to $s$ gives
\begin{equation}
g_{\mu\nu}\Bigl(\frac{du^\mu}{ds}+\{^{\,\,\mu}_{\rho\,\sigma}\}u^\rho u^\sigma\Bigr)u^\nu=0.
\label{geo}
\end{equation}
This equation is satisfied by
\begin{equation}
g_{\mu\nu}\Bigl(\frac{du^\mu}{ds}+\{^{\,\,\mu}_{\rho\,\sigma}\}u^\rho u^\sigma\Bigr)=a_{\nu\rho}u^\rho,
\label{Lor}
\end{equation}
where $a_{\nu\rho}$ is an arbitrary skewsymmetric tensor.
The simplest geometrical choice for $a_{\mu\nu}$ is to identify it with the electromagnetic field tensor $F_{\mu\nu}$ (which is a part of the affine connection~(\ref{con})):
\begin{equation}
a_{\mu\nu}=bF_{\mu\nu},
\label{prop}
\end{equation}
where $b$ is a physical constant.
If $b=\frac{e}{m}$, we obtain the classical Lorentz equation for a particle with mass $m$ and electric charge $e$.

\section{Metric-affine Gravity}

A general metric-affine Lagrangian density ${\cal L}$ depends on the affine connection $\Gamma^{\,\,\rho}_{\mu\,\nu}$ and the curvature tensor, $R^\rho_{\phantom{\rho}\mu\sigma\nu}$, as well as the metric tensor $g_{\mu\nu}$ with the Lorentzian signature $(+,-,-,-)$.
The simplest metric-affine Lagrangian density that depends on the curvature is the Einstein-Hilbert Lagrangian density for the gravitational field~\cite{MAE1,MAE2}:
\begin{equation}
{\cal L}_g=-\frac{1}{2\kappa}R_{\mu\nu}{\sf g}^{\mu\nu},
\label{LagGrav}
\end{equation}
where $\kappa=8\pi G$ ($c=1$), ${\sf g}^{\mu\nu}=\sqrt{-g}g^{\mu\nu}$ is the fundamental metric density~\cite{Schr}, and $g=\mbox{det}(g_{\mu\nu})$.
The total Lagrangian density for the gravitational field and of matter is given by ${\cal L}={\cal L}_g+{\cal L}_m$, where the Lagrangian density for matter ${\cal L}_m$ depends in general on both the metric and connection.
The variations of ${\cal L}_m$ with respect to the metric and connection define, respectively, the dynamical energy-momentum tensor:
\begin{equation}
T_{\mu\nu}=\frac{2}{\sqrt{-g}}\frac{\delta{\cal L}_m}{\delta g^{\mu\nu}},
\label{emt}
\end{equation}
and the hypermomentum density~\cite{MA3,MA4,MA5,MA6}:
\begin{equation}
\Pi_{\phantom{\mu}\rho\phantom{\nu}}^{\mu\phantom{\rho}\nu}=-2\kappa\frac{\delta{\cal L}_m}{\delta \Gamma^{\,\,\rho}_{\mu\,\nu}},
\label{Pi}
\end{equation}
which has the same dimension as the connection.\footnote{
The variational derivative of a function ${\cal L}(\phi,\phi_{,\mu})$ with respect to a variable $\phi$ is defined as $\frac{\delta{\cal L}}{\delta\phi}=\frac{\partial{\cal L}}{\partial\phi}-(\frac{\partial{\cal L}}{\partial\phi_{,\mu}})_{,\mu}$.
}

The metric-affine theory based on the Lagrangian density ${\cal L}_g$ does not determine the connection uniquely because ${\cal L}_g$~(\ref{LagGrav}) depends only on the symmetric part $R_{(\mu\nu)}$ of the Ricci tensor and thus is invariant under projective transformations~\cite{PO,MA3,MA4,MA5,MA6}:
\begin{equation}
\Gamma^{\,\,\rho}_{\mu\,\nu}\rightarrow\Gamma^{\,\,\rho}_{\mu\,\nu}+\delta^\rho_\mu V_\nu,
\label{proj}
\end{equation}
where $V_\nu$ is a vector function of the coordinates.
The same problem occurs if we add to ${\cal L}_g$ a Lagrangian for matter ${\cal L}_m$ that does not depend on the connection, for example, representing the electromagnetic field or an ideal fluid.
Therefore at least four degrees of freedom must be constrained to make such a theory consistent from a physical point of view~\cite{MA3,MA4,MA5,MA6}.
If ${\cal L}_m$ does depend on the connection, for example, for spinor fields, the projective invariance of the total Lagrangian density ${\cal L}={\cal L}_g+{\cal L}_m$ imposes four algebraic constraints on $\Pi_{\phantom{\mu}\rho\phantom{\nu}}^{\mu\phantom{\rho}\nu}$ (cf. Eq.~(\ref{cons})) and restricts forms of matter that can be described by metric-affine gravity~\cite{MA3,MA4,MA5,MA6}.
This restriction has usually the form of a field equation.
For example, including the term proportional to $\sqrt{-g}Q_{\mu\nu}Q^{\mu\nu}$ which has the form of the Maxwell Lagrangian for the electromagnetic field, gives the Maxwell-like equations for the tensor $Q_{\mu\nu}$~\cite{PO}.

From the stationarity of the action $S=\int d^4x{\cal L}$ under arbitrary variations of $g^{\mu\nu}$: $\delta S=0$, we obtain the metric-affine Einstein equations:
\begin{equation}
R_{(\mu\nu)}-\frac{1}{2}Rg_{\mu\nu}=\kappa T_{\mu\nu}.
\label{Eins}
\end{equation}
The variation of the action for the Lagrangian density~(\ref{LagGrav}) with respect to the connection is: $\delta S_g=-\frac{1}{2\kappa}\int d^4x\,{\sf g}^{\mu\nu}\delta R_{\mu\nu}$.
Using the Palatini formula for the variation of the Ricci tensor~\cite{Schr,Scho}: $\delta R_{\mu\nu}=\delta\Gamma^{\,\,\rho}_{\mu\,\nu;\rho}-\delta\Gamma^{\,\,\rho}_{\mu\,\rho;\nu}-2S^\sigma_{\phantom{\sigma}\rho\nu}\delta\Gamma^{\,\,\rho}_{\mu\,\sigma}$, we find
\begin{equation}
\delta S=-\frac{1}{2\kappa}\int d^4x\Bigl({\sf g}^{\mu\nu}(\delta\Gamma^{\,\,\rho}_{\mu\,\nu;\rho}-\delta\Gamma^{\,\,\rho}_{\mu\,\rho;\nu}-2S^\sigma_{\phantom{\sigma}\rho\nu}\delta\Gamma^{\,\,\rho}_{\mu\,\sigma})+\Pi_{\phantom{\mu}\rho\phantom{\nu}}^{\mu\phantom{\rho}\nu}\delta \Gamma^{\,\,\rho}_{\mu\,\nu}\Bigr).
\label{var}
\end{equation}
Integrating by parts and using the identity $\int d^4x({\sf V}^\mu)_{;\mu}=2\int d^4x S_\mu{\sf V}^\mu$, where ${\sf V}^\mu$ is an arbitrary vector density, from the stationarity of the action under arbitrary variations of $\Gamma^{\,\,\rho}_{\mu\,\nu}$ we obtain
\begin{equation}
{\sf g}^{\mu\nu}_{\phantom{\mu\nu};\rho}-{\sf g}^{\mu\sigma}_{\phantom{\mu\sigma};\sigma}\delta^\nu_\rho-2{\sf g}^{\mu\nu}S_\rho+2{\sf g}^{\mu\sigma}S_\sigma\delta^\nu_\rho+2{\sf g}^{\mu\sigma}S^\nu_{\phantom{\nu}\rho\sigma}=\Pi_{\phantom{\mu}\rho\phantom{\nu}}^{\mu\phantom{\rho}\nu}.
\label{field1}
\end{equation}
Contracting the indices $\mu$ and $\rho$ in Eq.~(\ref{field1}) gives
\begin{equation}
\Pi_{\phantom{\sigma}\sigma\phantom{\nu}}^{\sigma\phantom{\sigma}\nu}=0,
\label{cons}
\end{equation}
which constrains how the connection $\Gamma^{\,\,\rho}_{\mu\,\nu}$ can enter the metric-affine Lagrangian density for matter ${\cal L}_m$.
Equation~(\ref{field1}) is equivalent to
\begin{equation}
{\sf g}^{\mu\nu}_{\phantom{\mu\nu},\rho}+\,^\ast\Gamma^{\,\,\mu}_{\sigma\,\rho}{\sf g}^{\sigma\nu}+\,^\ast\Gamma^{\,\,\nu}_{\rho\,\sigma}{\sf g}^{\mu\sigma}-\,^\ast\Gamma^{\,\,\sigma}_{\sigma\,\rho}{\sf g}^{\mu\nu}=\Pi_{\phantom{\mu}\rho\phantom{\nu}}^{\mu\phantom{\rho}\nu}-\frac{1}{3}\Pi_{\phantom{\mu}\sigma\phantom{\sigma}}^{\mu\phantom{\sigma}\sigma}\delta^\nu_\rho,
\label{field2}
\end{equation}
where $^\ast\Gamma^{\,\,\rho}_{\mu\,\nu}=\Gamma^{\,\,\rho}_{\mu\,\nu}+\frac{2}{3}\delta^\rho_\mu S_\nu$ is the projectively invariant part of the connection (Schr\"{o}dinger's star-affinity)~\cite{Schr}.

Equation~(\ref{field2}) gives a linear relation between $^\ast\Gamma^{\,\,\rho}_{\mu\,\nu}$ and the hypermomentum density $\Pi_{\phantom{\mu}\rho\phantom{\nu}}^{\mu\phantom{\rho}\nu}$.
If we decompose the star-affinity $^\ast\Gamma^{\,\,\rho}_{\mu\,\nu}$ as
\begin{equation}
^\ast\Gamma^{\,\,\rho}_{\mu\,\nu}=\{^{\,\,\rho}_{\mu\,\nu}\}+V^\rho_{\phantom{\rho}\mu\nu},
\label{dec}
\end{equation}
where $V^\rho_{\phantom{\rho}\mu\nu}$ is a projectively invariant deflection tensor, then $V^\rho_{\phantom{\rho}\mu\nu}$ is linear in $\Pi_{\phantom{\mu}\rho\phantom{\nu}}^{\mu\phantom{\rho}\nu}$:
\begin{equation}
V^\mu_{\phantom{\mu}\sigma\rho}{\sf g}^{\sigma\nu}+V^\nu_{\phantom{\nu}\rho\sigma}{\sf g}^{\mu\sigma}-V^\sigma_{\phantom{\sigma}\sigma\rho}{\sf g}^{\mu\nu}=\Pi_{\phantom{\mu}\rho\phantom{\nu}}^{\mu\phantom{\rho}\nu}-\frac{1}{3}\Pi_{\phantom{\mu}\sigma\phantom{\sigma}}^{\mu\phantom{\sigma}\sigma}\delta^\nu_\rho.
\label{field3}
\end{equation}
The solution of Eq.~(\ref{field3}) is given by (cf. Ref.~\cite{NikoA})
\begin{eqnarray}
& & V^\rho_{\phantom{\rho}\mu\nu}=\frac{1}{2\sqrt{-g}}(\Delta_{\phantom{\rho}\nu\phantom{\sigma}}^{\rho\phantom{\nu}\sigma}g_{\mu\sigma}+\Delta_{\phantom{\rho}\mu\phantom{\sigma}}^{\rho\phantom{\mu}\sigma}g_{\nu\sigma}-\Delta_{\phantom{\alpha}\gamma\phantom{\beta}}^{\alpha\phantom{\gamma}\beta}g_{\mu\alpha}g_{\nu\beta}g^{\rho\gamma} \nonumber \\
& & +\Omega_\nu^{\phantom{\nu}\rho\sigma}g_{\mu\sigma}-\Omega_\mu^{\phantom{\mu}\rho\sigma}g_{\nu\sigma}-\Omega_\gamma^{\phantom{\gamma}\alpha\beta}g_{\mu\alpha}g_{\nu\beta}g^{\rho\gamma}),
\label{sol}
\end{eqnarray}
where
\begin{eqnarray}
& & \Delta_{\phantom{\mu}\rho\phantom{\nu}}^{\mu\phantom{\rho}\nu}=\Sigma_{\phantom{\mu}\rho\phantom{\nu}}^{\mu\phantom{\rho}\nu}-\frac{1}{2}\Sigma_{\phantom{\alpha}\rho\phantom{\beta}}^{\alpha\phantom{\rho}\beta}g_{\alpha\beta}g^{\mu\nu}, \\
\label{delta}
& & \Sigma_{\phantom{\mu}\rho\phantom{\nu}}^{\mu\phantom{\rho}\nu}=\Pi_{\phantom{(\mu}\rho\phantom{\nu)}}^{(\mu\phantom{\rho}\nu)}-\frac{1}{3}\delta^{(\mu}_\rho\Pi_{\phantom{\nu)}\sigma\phantom{\sigma}}^{\nu)\phantom{\sigma}\sigma}, \\
& & \Omega_\rho^{\phantom{\rho}\mu\nu}=\Pi_{\phantom{[\mu}\rho\phantom{\nu]}}^{[\mu\phantom{\rho}\nu]}+\frac{1}{3}\delta^{[\mu}_\rho\Pi_{\phantom{\nu]}\sigma\phantom{\sigma}}^{\nu]\phantom{\sigma}\sigma}.
\label{omega}
\end{eqnarray}
For the connection given by Eq.~(\ref{dec}), the Ricci tensor is quadratic in $V^\rho_{\phantom{\rho}\mu\nu}$~\cite{Scho}, that is, in $\Pi_{\phantom{\mu}\rho\phantom{\nu}}^{\mu\phantom{\rho}\nu}$:
\begin{equation}
R_{\mu\nu}=K_{\mu\nu}-\frac{4}{3}S_{[\nu:\mu]}+2V^\rho_{\phantom{\rho}\mu[\nu:\rho]}+V^\sigma_{\phantom{\sigma}\mu\nu}V^\rho_{\phantom{\rho}\sigma\rho}-V^\sigma_{\phantom{\sigma}\mu\rho}V^\rho_{\phantom{\rho}\sigma\nu},
\label{Ein1}
\end{equation}
Substituting this $R_{\mu\nu}$ to the metric-affine Einstein equations~(\ref{Eins}) and moving the terms with $V^\rho_{\phantom{\rho}\mu\nu}$ to the right-hand side gives the metric Einstein equations.
The Ricci scalar is~\cite{Scho}
\begin{equation}
R=K+V^{\rho\sigma}_{\phantom{\rho\sigma}\sigma:\rho}-V_{\rho\phantom{\rho\sigma}:\sigma}^{\phantom{\rho}\rho\sigma}+V^{\sigma\lambda}_{\phantom{\sigma\lambda}\lambda}V^\rho_{\phantom{\rho}\sigma\rho}-V_{\sigma\lambda\rho}V^{\rho\sigma\lambda}.
\label{Ein2}
\end{equation}

\section{Variational Formulation of Eisenhart's Theory}

Comparing Eq.~(\ref{con}) with~(\ref{dec}) and using the condition~(\ref{Max}) gives
\begin{equation}
V^\rho_{\phantom{\rho}\mu\nu}=kF_{\mu\nu:\sigma}g^{\rho\sigma}.
\label{defl1}
\end{equation}
Equation~(\ref{field3}) gives
\begin{equation}
\Pi_{\phantom{\mu}\rho\phantom{\nu}}^{\mu\phantom{\rho}\nu}=\sqrt{-g}(V^{\mu\nu}_{\phantom{\mu\nu}\rho}+V^{\nu\phantom{\rho}\mu}_{\phantom{\nu}\rho}-V^\sigma_{\phantom{\sigma}\sigma\rho}g^{\mu\nu}-V^{\mu\phantom{\sigma}\sigma}_{\phantom{\mu}\sigma}\delta^\nu_\rho).
\label{defl2}
\end{equation}
Substituting Eq.~(\ref{defl1}) to~(\ref{defl2}) gives the hypermomentum density:
\begin{equation}
\Pi_{\phantom{\mu}\rho\phantom{\nu}}^{\mu\phantom{\rho}\nu}=k\sqrt{-g}g^{\mu\alpha}g^{\nu\beta}(F_{\rho\alpha:\beta}-F_{\rho\beta:\alpha}),
\label{defl3}
\end{equation}
from which it follows that the matter Lagrangian density ${\cal L}_m$ that generates~(\ref{defl3}) via Eq.~(\ref{Pi}) is given by\footnote{
We use the fact that the variation with respect to $\Gamma^{\,\,\rho}_{\mu\,\nu}$ in metric-affine gravity is equivalent to the variation with respect to $\Gamma^{\,\,\rho}_{\mu\,\nu}-\{^{\,\,\rho}_{\mu\,\nu}\}$ and the covariant derivative $:$ does not depend on $\Gamma^{\,\,\rho}_{\mu\,\nu}-\{^{\,\,\rho}_{\mu\,\nu}\}$.
\label{foot}
}
\begin{equation}
{\cal L}_m=-\frac{k}{\kappa}\sqrt{-g}F_{\rho\alpha:\beta}S^\rho_{\phantom{\rho}\mu\nu}g^{\mu\alpha}g^{\nu\beta}.
\label{Lagm}
\end{equation}
This Lagrangian density, however, also generates the energy-momentum tensor via Eq.~(\ref{emt}):
\begin{equation}
T_{\mu\nu}=\frac{k}{\kappa}(F_{\rho\alpha:\beta}S^{\rho\alpha\beta}g_{\mu\nu}-2F_{\rho[\alpha:\mu]}S^{\rho\alpha}_{\phantom{\rho\alpha}\nu}-2F_{\rho[\alpha:\nu]}S^{\rho\alpha}_{\phantom{\rho\alpha}\mu}),
\label{add}
\end{equation}
which is traceless.
The total Lagrangian density is equal to\footnote{
The Lagrangian density~(\ref{Lagr}) is dynamically equivalent to ${\cal L}=\sqrt{-g}\Bigl(-\frac{1}{2\kappa}R+\frac{k}{\kappa}F_{\rho\alpha}S^{\rho\alpha\beta}_{\phantom{\rho\alpha\beta}:\beta}\Bigr)$, in which the electromagnetic field tensor couples to the divergence of the torsion tensor $S^{\mu\nu\rho}_{\phantom{\mu\nu\rho}:\rho}$.
}
\begin{equation}
{\cal L}=\sqrt{-g}\Bigl(-\frac{1}{2\kappa}R-\frac{k}{\kappa}F_{\rho\alpha:\beta}S^{\rho\alpha\beta}\Bigr),
\label{Lagr}
\end{equation}
and produces the field equations~(\ref{Eins}) which, using Eqs.~(\ref{tor}) and~(\ref{add}), can be written as
\begin{equation}
R_{(\mu\nu)}=k^2(F^\rho_{\phantom{\rho}\alpha:\beta}F^{\alpha\beta}_{\phantom{\alpha\beta}:\rho}g_{\mu\nu}-2F^\rho_{\phantom{\rho}[\alpha:\mu]}F^\alpha_{\phantom{\alpha}\nu:\rho}-2F^\rho_{\phantom{\rho}[\alpha:\nu]}F^\alpha_{\phantom{\alpha}\mu:\rho}),
\label{field4}
\end{equation}
or, using Eq.~(\ref{Ric}), as
\begin{equation}
K_{\mu\nu}=k^2(F^\rho_{\phantom{\rho}\alpha:\beta}F^{\alpha\beta}_{\phantom{\alpha\beta}:\rho}g_{\mu\nu}-2F^\rho_{\phantom{\rho}[\alpha:\mu]}F^\alpha_{\phantom{\alpha}\nu:\rho}-2F^\rho_{\phantom{\rho}[\alpha:\nu]}F^\alpha_{\phantom{\alpha}\mu:\rho}-F_{\mu\phantom{\rho}:\sigma}^{\phantom{\mu}\rho}F_{\nu\phantom{\sigma}:\rho}^{\phantom{\nu}\sigma}).
\label{field6}
\end{equation}

We can obtain a simpler field equation if we add to the right-hand side of Eq.~(\ref{Lagr}) a term: $\tilde{{\cal L}}_m=\frac{k^2}{\kappa}F^\rho_{\phantom{\rho}\alpha:\beta}F^{\alpha\beta}_{\phantom{\alpha\beta}:\rho}$, so the total Lagrangian density is
\begin{equation}
{\cal L}=\sqrt{-g}\Bigl(-\frac{1}{2\kappa}R-\frac{k}{\kappa}F_{\rho\alpha:\beta}S^{\rho\alpha\beta}+\frac{k^2}{\kappa}F^\rho_{\phantom{\rho}\alpha:\beta}F^{\alpha\beta}_{\phantom{\alpha\beta}:\rho}\Bigr).
\label{ff}
\end{equation}
Equation~(\ref{defl3}) does not change (cf. footnote~\ref{foot}), while the variation of the action corresponding to the Lagrangian density~(\ref{ff}) with respect to $g_{\mu\nu}$, together with Eq.~(\ref{tor}), gives
\begin{equation}
R_{(\mu\nu)}=k^2(2F^{\alpha\beta}_{\phantom{\alpha\beta}:(\mu}F_{\nu)\alpha:\beta}-F^\rho_{\phantom{\rho}\alpha:\beta}F^{\alpha\beta}_{\phantom{\alpha\beta}:\rho}g_{\mu\nu}),
\label{field5}
\end{equation}
or
\begin{equation}
K_{\mu\nu}=k^2(2F^{\alpha\beta}_{\phantom{\alpha\beta}:(\mu}F_{\nu)\alpha:\beta}-F^\rho_{\phantom{\rho}\alpha:\beta}F^{\alpha\beta}_{\phantom{\alpha\beta}:\rho}g_{\mu\nu}-F_{\mu\phantom{\rho}:\sigma}^{\phantom{\mu}\rho}F_{\nu\phantom{\sigma}:\rho}^{\phantom{\nu}\sigma}).
\label{field7}
\end{equation}
In order to reproduce Eisenhart's equation~(\ref{Eis}), we should add to the right-hand side of Eq.~(\ref{Lagr}) a term that does not depend on $\Gamma^{\,\,\rho}_{\mu\,\nu}-\{^{\,\,\rho}_{\mu\,\nu}\}$ and that generates, via Eq.~(\ref{emt}), the energy-momentum tensor $-\frac{k^2}{\kappa}(F^\rho_{\phantom{\rho}\alpha:\beta}F^{\alpha\beta}_{\phantom{\alpha\beta}:\rho}g_{\mu\nu}-2F^\rho_{\phantom{\rho}[\alpha:\mu]}F^\alpha_{\phantom{\alpha}\nu:\rho}-2F^\rho_{\phantom{\rho}[\alpha:\nu]}F^\alpha_{\phantom{\alpha}\mu:\rho})$ which cancels the energy-momentum tensor~(\ref{add}).

The problem with Eisenhart's field equation~(\ref{Eis}) and its modifications~(\ref{field6}) and~(\ref{field7}) occurs when we apply to them the contracted Bianchi identity $(K_\mu^{\phantom{\mu}\nu}-\frac{1}{2}K\delta_\mu^\nu)_{:\nu}=0$.
This identity gives additional constraints on the electromagnetic field tensor, which is inconsistent with the Maxwell electrodynamics.
Therefore we must add to the right-hand side of Eq.~(\ref{Lagr}) a term that does not depend on $\Gamma^{\,\,\rho}_{\mu\,\nu}-\{^{\,\,\rho}_{\mu\,\nu}\}$ and that generates, via Eq.~(\ref{emt}), the energy-momentum tensor $-\frac{k^2}{\kappa}(F^\rho_{\phantom{\rho}\alpha:\beta}F^{\alpha\beta}_{\phantom{\alpha\beta}:\rho}g_{\mu\nu}-2F^\rho_{\phantom{\rho}[\alpha:\mu]}F^\alpha_{\phantom{\alpha}\nu:\rho}-2F^\rho_{\phantom{\rho}[\alpha:\nu]}F^\alpha_{\phantom{\alpha}\mu:\rho}-F_{\mu\phantom{\rho}:\sigma}^{\phantom{\mu}\rho}F_{\nu\phantom{\sigma}:\rho}^{\phantom{\nu}\sigma})$, resulting in Einstein's field equation $K_{\mu\nu}=0$.

The above variational formulation of Eisenhart's theory is not complete because we did not include the condition that the torsion vector $S_\mu$ be zero in the Lagrangian.
This condition enters the Lagrangian density as a Lagrange-multiplier term $\sqrt{-g}B^\mu S_\mu$, where $B^\mu$ is a vector~\cite{MA2,NikoA}.
The total Lagrangian density~(\ref{Lagr}) becomes
\begin{equation}
{\cal L}=\sqrt{-g}\Bigl(-\frac{1}{2\kappa}R-\frac{k}{\kappa}F_{\rho\alpha:\beta}S^{\rho\alpha\beta}+B^\mu S_\mu\Bigr).
\label{mult1}
\end{equation}
Treating the term $\sqrt{-g}B^\mu S_\mu$ as a matter part of the Lagrangian adds, via Eq.~(\ref{Pi}), a term $-2\kappa\sqrt{-g}B^{[\mu}\delta^{\nu]}_\rho$ to the hypermomentum density $\Pi_{\phantom{\mu}\rho\phantom{\nu}}^{\mu\phantom{\rho}\nu}$.
Since the deflection tensor $V^\rho_{\phantom{\rho}\mu\nu}$ is linear in $\Pi_{\phantom{\mu}\rho\phantom{\nu}}^{\mu\phantom{\rho}\nu}$, the corresponding change in $V^\rho_{\phantom{\rho}\mu\nu}$ is the solution of Eq.~(\ref{sol}) for $\Pi_{\phantom{\mu}\rho\phantom{\nu}}^{\mu\phantom{\rho}\nu}=-2\kappa\sqrt{-g}B^{[\mu}\delta^{\nu]}_\rho$, that is
\begin{equation}
V^\rho_{\phantom{\rho}\mu\nu}=\frac{\kappa}{4}(3B^\rho g_{\mu\nu}+B_\mu\delta^\rho_\nu-3B_\nu\delta^\rho_\mu).
\label{mult2}
\end{equation}
The total deflection tensor is
\begin{equation}
V^\rho_{\phantom{\rho}\mu\nu}=kF_{\mu\nu:\sigma}g^{\rho\sigma}+\frac{\kappa}{4}(3B^\rho g_{\mu\nu}+B_\mu\delta^\rho_\nu-3B_\nu\delta^\rho_\mu),
\label{mult3}
\end{equation}
which gives, using Eq.~(\ref{dec}), the torsion vector:
\begin{equation}
S_\mu=-kF^\nu_{\phantom{\nu}\mu:\nu}+\frac{3\kappa}{2}B_\mu.
\label{mult4}
\end{equation}
Varying the action corresponding to~(\ref{mult1}) with respect to $B^\mu$ gives $S_\mu=0$ which results in
\begin{equation}
B^\mu=\frac{2k}{3\kappa}j^\mu,
\label{mult5}
\end{equation}
where $j^\mu=F^{\nu\mu}_{\phantom{\nu\mu}:\nu}$ is the conserved ($j^\mu_{\phantom{\mu}:\mu}=0$) electromagnetic current vector.
Therefore the variational formulation of Eisenhart's unified theory generalizes this theory to electromagnetic fields with sources represented by a Lagrange-multiplier constraint that imposes a traceless torsion.

\section{Concluding Remarks}

Classical unified field theory is a topic worthy of investigation because it can be regarded as a classical limit of a quantum unified theory.
Because we already have a successful quantum theory of the electromagnetic field (QED), knowing how to combine gravity with electromagnetic interaction (the only interaction beside gravitation that is significant at the classical limit) can give us insights on how to quantize the gravitational field.
In this paper, we formulated Eisenhart's classical unified theory of gravitation and electromagnetism~\cite{Eis1} as a variational theory.
We used the metric-affine formulation of gravity and showed that the variational formulation of Eisenhart's theory generalizes this theory to electromagnetic fields with sources.
This mathematical exercise can thus serve as an example of how sources of a physical field arise in a theory based on a variational principle.

Equation~(\ref{add}) for the energy-momentum tensor of the electromagnetic field is unphysical because the corresponding energy density of this field depends on the derivatives of the electromagnetic field tensor $F_{\mu\nu}$ instead of $F_{\mu\nu}$ itself.
If we associate the tensor $F_{\mu\nu}$ with the non-Riemannian part of the affine connection, as in Eisenhart's theory, then the Ricci tensor contains terms that are quadratic in $F_{\mu\nu}$, as in the Einstein-Maxwell theory.
However, the connection has three indices while $F_{\mu\nu}$ has only two, so such an association requires an additional index related to some external field.
On the other hand, if we associate the electromagnetic field potential $A_\mu$ with the non-Riemannian part of the affine connection, as in Refs.~\cite{FKu} and~\cite{PO}, the resulting unified theory contains a correct expression for the energy-momentum tensor of the electromagnetic field and is dynamically equivalent to the Einstein-Maxwell theory with sources, without using Lagrange multipliers in the action.

\end{document}